# Stern-Gerlach: conceptually clean or acceptably vague?

*Robert Shaw (shaw2812@gmail.com)*

This paper develops a number of quantum mechanical characterisations of Stern-Gerlach. It discusses areas of vagueness in their formulation. Philosophers criticise quantum mechanics for unacceptable vagueness in connection with the measurement problem. The quantum formulation problems identified by this paper go beyond the locus of philosophical criticism. It concludes with an open question, are some areas of vagueness in quantum mechanics more acceptable philosophically than others and, if so, why?

I. **INTRODUCTION**

"Stern-Gerlach has played an important role in the philosophy of quantum mechanics where it serves as the prototype…it is conceptually very clean", according to Griffiths[1] in his *Introduction to Quantum Mechanics*.

Maudlin, in *Philosophy of Physics: Quantum Theory*, says,[2] "Quantum theory is a recipe or prescription using somewhat vague terms." and elaborates,[3] "the quantum recipe has three distinct parts" which can be summarised in fundamental terms as:

1. **State rules**: how a state is assigned to a system
2. **Transition rules**: how the state changes as a result of interactions (Schrodinger's equation)
3. **Measurement rules**: how predictions are extracted about observable phenomena

Part 3 is unacceptably vague for philosophers,[4] "what we want instead is a theory – a precise articulation" and this is the central problem of significance to them,[5] "often referred to as the measurement problem".

Using the example of the orthodox treatment of the Stern-Gerlach experiment, he criticises the simplification and lack of rigor, saying it,[6] "short circuits all the gritty physics" and what's required is, "a physical characterisation of the situation."

This paper attempts such a gritty physical characterisation of Stern-Gerlach. In doing so it focuses on rules 1 and 2, given that rule 3 has been extensively studies by philosophers. It discusses whether Rules 1 and 2 are also vague. It concludes with the open question, where vagueness is encountered for rules 1 and 2 why do the quantum philosophers neglect to discuss the issue?

II. **A SIMPLIFIED QUANTUM MECHANICAL CALCULATION**

This section sets out a simplified calculation, a toy model that misses out many of the details, which helps develop the structure and logic. Even a simplified calculation is challenging. Platt[7] comments that "While the Stern-Gerlach experiment is an old and familiar problem, no analysis is presented in the pedagogical literature using modern quantum mechanical techniques." He and several other authors[8 9 10 11 12] report various ways in which the atom-magnet interaction can be calculated using quantum mechanics. These calculations all differ from one another in materially significant ways.

The common ingredients of a simplified quantum mechanical calculation are: a state-space for the "silver atom"; a formula for the silver atom's Hamiltonian; an initial state for the silver atom; time between the atom entering and exiting the magnet; time between the atom exiting the magnet and arriving at the detector. An example of a calculation using these ingredients is set out below.

A likely structure for the state-space can be guessed. Position is one observable and its eigenstates are labelled by x, y and z co-ordinates; alternatively, a momentum representation can be used interchangeably with position, to facilitate with the calculations.

The other observable operator is the magnetic moment $\hat{\mu}$ of silver. Its eigenstates will be assumed to take on two values, represented by a Pauli spinor, and $\hat{\mu} = \mu \, \hat{\sigma}_z$. The structure of a state vector on the momentum basis is:

$$|\Psi(t)\rangle = \begin{vmatrix} U(p_x, p_y, p_z, t) \\ D(p_x, p_y, p_z, t) \end{vmatrix}$$

The field inside the Stern-Gerlach magnet has the form (Platt[13]):

$$B_x = 0 \; ; \; B_y = \beta y \; ; \; B_z = B_0 - \beta \hat{z}$$

The Hamiltonian can be constructed using classical mechanics and it has the form:

$$\hat{H} = \hat{p}^2/2M_{Ag} + \mu_{Ag} [\hat{\sigma}_x B_x(x,y,z) + \hat{\sigma}_y B_y(x,y,z) + \hat{\sigma}_z B_z(x,y,z)]$$

where the constant $M_{Ag}$ is the mass of the silver atom and $\mu_{Ag}$ is its magnetic moment. A further approximation is useful to facilitate an analytical solution - only the $B_z$ component is retained. The equations of motion then become separable and U and D have their own Hamiltonians:

$$\hat{H}_U = \hat{p}^2/2M_{Ag} + \mu_{Ag}(B_0 - \beta \hat{z})$$

$$\hat{H}_D = \hat{p}^2/2M_{Ag} - \mu_{Ag}(B_0 - \beta \hat{z})$$

We now make use of the commutation relations:

$$[\hat{z}, \hat{p}_z] = i\hbar$$

and hence:

$$[\hat{H}, \hat{p}_x] = [\hat{H}, \hat{p}_y] = 0$$

$$[\hat{H}_U, \hat{p}_z] = -\mu_{Ag} \beta \, i\hbar$$

$$[\hat{H}_D, \hat{p}_z] = +\mu_{Ag} \beta \, i\hbar$$

from which it is concluded that $\hat{p}_x$ and $\hat{p}_y$ are constants of motion, and $\hat{p}_z$ evolves while inside the magnet.

The following assumptions about initial conditions are made:

- initial position of the wavepacket is $\langle x \rangle = \langle y \rangle = \langle z \rangle = 0$ (at the entry point of the magnet)
- initial momentum of the wavepacket is $\langle \hat{p}_x \rangle = \langle \hat{p}_z \rangle = 0$; $\langle \hat{p}_y \rangle = M_{Ag} V_y$
- spread of the wavepacket in space and momentum is macroscopically small

For the U component, motion inside the magnet is calculated as follows:

$$d\langle \hat{p}_z \rangle/dt = -\mu_{Ag} \beta$$

$$d\langle \hat{z} \rangle/dt = \langle \hat{p}_z \rangle/M_{Ag}$$

$$d\langle \hat{y} \rangle/dt = \langle \hat{p}_y \rangle/M_{Ag} = V_y$$

$$\langle \hat{y} \rangle = V_y t$$

$$\langle \hat{z} \rangle = -\mu_{Ag} \beta t^2 / 2 M_{Ag} = -\langle \hat{y} \rangle^2 \mu_{Ag} \beta / 2 V_y^2 M_{Ag}$$

$$\mu_{Ag} = -\langle \hat{z} \rangle 2 V_y^2 M_{Ag} / \langle \hat{y} \rangle^2 \beta$$

The interaction Hamiltonian responsible for the motion is consists of three mathematical terms: the displacement of the silver, the magnetic moment of the silver and the field gradient of the magnet.

$$\hat{H}_{int} = -\hat{z}_{Ag} \hat{\mu}_{Ag} \beta$$

Wigner[14] interprets the situation as follows: "The experiment illustrates the statistical correlation between the state of the apparatus (the position coordinate) and the state of the object (the spin)." Wigner's logic is shown schematically in Figure 2.

Figure 2: Wigner's view of the Stern-Gerlach interaction

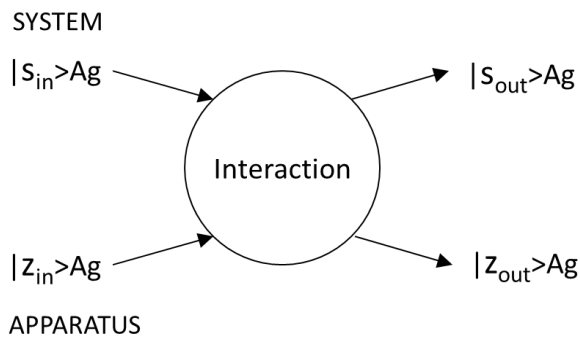

This raises a question as to what exactly is an interaction? Is Wigner arguing that the presence in the Hamiltonian of operators for position and magnetic moment of the silver mean that the silver is "interacting" with itself?

An alternative view would be to include the magnet in the interaction, as illustrated in Figure 3. This formulation of the interaction would enable an alternative method of measurement. If the magnet were to be measured after the interaction, for example by an instrument sensitive to forces at the atomic scale, then the deflection of the magnet |z_out⟩ could be used to extract information about the state of the silver atom before the interaction.

Figure 3: Interaction between silver and magnet

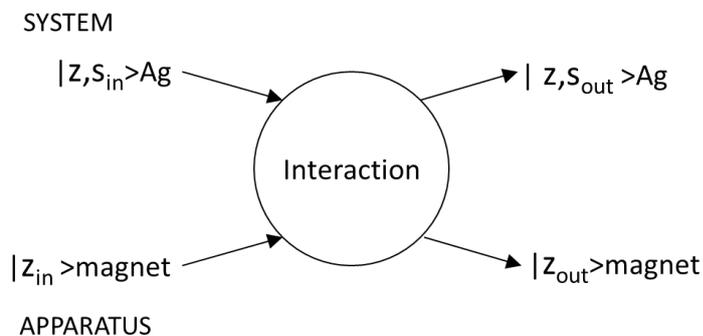

### III.  MAKING THE CALCULATIONS MORE COMPLETE

The previous calculations are an incomplete representation of the Stern-Gerlach experiment. This section considers a range of factors that may need to be considered to make the simplified calculation in section III more complete and, in so doing, increase its accuracy.

An atom is not a particle, but a composite object. The first refinement is to model it as 47 electrons orbiting a nucleus. The electrons arrange themselves into a shell structure: 2 8 18 18 1. The orbital angular momentum is 0. The nucleus and each of the electrons has spin ½; the total angular momentum of the atom in its ground state is therefore 0 (with the valence electron spin opposite to the nuclear spin) or 1 (with spins aligned). The magnetic moment of the valence electron $5s^1$ couples with the Stern-Gerlach magnet; all other electrons have contributions that cancel. So the experiment can be used to measure the spin of the valence electron but it does not enable the angular momentum of the atom itself to be measured.

The second refinement is to model the structure of the nucleus. Take the nucleus of the isotope $^{107}_{47}$Ag which is a composite of 47 protons and 60 neutrons. They are bound together by the residual strong force, and its description is largely empirical, with constants determined by experimental data. Because the nucleus has an odd number of particles, the magnetic moments of the nucleons do not cancel and there is a net magnetic moment for the nucleus. This nuclear magnetic moment couples with the Stern-Gerlach magnet and results in a small correction to the previous calculation.

The third refinement to the model is that the nucleons consist of quarks. The masses of the neutron and proton ($1.67 \cdot 10^{-27}$ kg) are dominated by the kinetic energies of the quarks which significantly exceed the quark masses (4.1 and $8.5 \cdot 10^{-30}$ kg for up and down quarks). These are the main contributors to the mass of the silver atom, which appears in the Hamiltonian in section 2.2.

The next correction concerns the fact that the actual experiment did not use just one atom but trillions or more. Collisions between the atoms could contribute to the deflection, if the vapour was sufficiently dense for the mean free path to be comparable to the dimensions of the apparatus, and the deflections so caused could potentially obscure the deflections caused by the magnet.

The final correction concerns the material used in the experiment, "silver", which is not one thing but a mixture. The element has two stable isotopes. Their masses differ and this would result in fine structure with four beams emerging from the Stern-Gerlach magnet rather than two.

### IV.  HOW VAGUE IS THE QUANTUM RECIPE?

Maudlin suggests that quantum mechanics is a recipe and that Part 3, the measurement rule, is unacceptably vague. This section discusses whether Parts 1 and 2 are also vague.

#### A.  Part 1: The State Rule

The state rule concerns how a state is assigned to a system. Nielsen and Huang comment[15] : "Quantum mechanics does *not* tell us, for a given system, what is the state space of that system is…Figuring that out for a specific system is a difficult problem for which physicists have developed many intricate and beautiful rules." So the question about vagueness applies to many rules, which are apparently intricate and beautiful.

A conceptually clean mathematical representation of a state, in Dirac notation, would be |state>, a vector in Hilbert space. The notation inside the bracket contains all the information and the only information that specifies the state vector.

# Stern-Gerlach: conceptually clean or acceptably vague?

Yet the descriptions of the states in Stern-Gerlach are not conceptually clean in this sense. For example, Wigner's state is |s> contains only spin information and no information about the magnetic moment of silver, which will be needed to formulate the Hamiltonian. It makes no mention of silver at all.

The representation in Figure 3 has a system state |z,s>. Alternatively, the magnetic moment could be substituted for spin |z,μ> and additionally the mass could be treated as an observable |z,μ,M>. Another alternative would be to write |s,Ag> and assume that Ag is an observable. Most chemists can offer a test for silver, so in that sense it is an observable. Then again it could be argued that the concept of Ag itself is vague; it has several isotopes.

Section III raises further questions, by going down into the fine structure of silver and going up into macroscopic representations of silver vapour. Going down to the most elementary components in the Standard Model the state would be |particle1, particle2…particle368> for the silver isotope 107. From this model, in theory, the masses and magnetic moments would emerge.

Going up to a macroscopic level, the temperature T of the vapour is $1000^oC$, and from this the mean velocity $<V_y>$ to be calculated. Should the temperature also be included |z,μ,M,T>? The only place where temperature occurs in the Standard Model is for the quark-gluon plasma, and not the temperature of the silver vapour.

A notation found in the literature is to append a subscript outside the bra and ket delimiters, for example $|s>_{Ag}$ to denote that the state is "silver". However, the symbol Ag is not part of the quantum recipe, it has no home in the Hilbert space.

In real applications the general notation might be something like $|state>_{vague\ stuff}$. The vague stuff is not represented in Hilbert space and might include "silver". This allows "physical constants" associated with the vague stuff to be included in the Hamiltonian, in addition to "fundamental constants". All of which is highly subjective. Different scholars would have different criteria for what is or isn't *vague stuff*.

In the quantum philosophy literature, vague stuff is embraced within the formulation of quantum states. Take the example of Schrodinger's cat, which according to the orthodox treatment has only two states: $|alive>_{cat}$ or $|dead>_{cat}$. Biologists cannot agree the meaning of dead or alive, but in this thought experiment it is considered acceptable to represent the cat in this way.

*Factorise-and-forget* is an approach to defining the state. The classic example is the hydrogen atom. For a two-body system the Hamiltonian splits in two sub-Hamiltonians, one for the centre of mass $\hat{z}$ and the other for the relative position $\hat{r}$. Factorising the hydrogen state involves:

|hydrogen> = |r> x |z>

The radial position is interesting as it enables the energy levels of the hydrogen atom to be calculated. The centre-of-mass equation is trivial and so forgotten. In the case of the silver atom the factorise-and-forget strategy would be as follows.

|z,s,Ag> ➔ |z,s> x |Ag other stuff> ➔ |z,s>

For a two-body system factorise-and-forget is reasonable, when the system is isolated, but for a many-body system in a potential field it is not. Consider the silver atom in a potential field that at atomic scale has enormous gradients. Consider the representation |z,s,Ag>. Each particle "inside" the silver is subjected to huge potential gradients. These might ionise the electrons or even split the nucleus and "silver" would cease to exist.

The judgement as to what constitutes "enormous potential gradients" is subjective. If such subjectivity constitutes unacceptable vagueness, the only acceptable approach would be to model the system at its most elementary level as 368 elementary particles or perhaps at a lower level still as Standard Model field excitations. Entanglement is another feature of quantum mechanics that would also suggest factorizability should not be casually assumed when formulating quantum calculations.

What addendum to Rule 1 would enable a philosopher to decide which of these alternative formulations is better or best or right? A widely published addendum to Rule 1 seems to be *omit nothing*. Here are some expert opinions. "In classical physics knowing the state of a system implies knowing everything that is necessary to predict the future of that system…knowing a quantum state means knowing as much as possible about how the system was prepared" writes Susskind[16]. "The state of a physical system is usually a well-defined collection of information about the system…should be as complete as possible…to determine the future behaviour of the state" according to Thaller[17]; "The system is completely described by the state vector" comment Nielsen and Huang[18].

Despite being told by experts that the quantum state omits nothing, in reality factorise-and forget is the orthodox method of formulating quantum calculations. Stern-Gerlach and indeed all quantum calculations rely upon factorise-and forget assumptions. Factorise-and-forget is normal too in the philosophical literature about quantum mechanics, where toy models forget real details with casual abandon. Subjectivity is the real issue here. The choice of how to factorise and what to forget is a subjective judgement by the person, physicist or philosopher, who formulates the calculation. And the act of forgetting directly contradicts the omit nothing addendum to Rule 1.

B.  <u>Part 2: The Transition Rule</u>

The transition rule concerns how the state changes as a result of interactions. To apply it requires an understanding of the nature of the interactions, the relevant physical constants, and the mathematical methods needed to model them.

The standard formulation is the Schrodinger equation, which uses the quantum analogue of Hamiltonian mechanics, and is a second order differential equation. Alternatively, there is the Dirac equation which is a first order differential equation. Wigner's phase space formulation is a different approach that is widely used in solving practical problems. Particle physicists use an S-matrix.

For many-particle systems the states are represented by occupation number states and the equations of motion involve Lagrangians, which are different from Schrodinger's approach. Differential equations are less important and path integrals dominate in quantum field theory.

The variational formulation provides a full picture describing any state and the way it changes; this approach is now the preferred recipe for studying supersymmetric strings or membranes. Quantum optics has extended again the methods available. In particular for studying coherent states it has its own range of techniques.

Hybrid approaches are also commonplace. Hawking radiation is described by a hybrid of quantum virtual photons moving under the rules of general relativity. Condensed matter physics brings together methods from field theory, statistical mechanics and discovers "emergent phenomena" that come to light from the equations in surprising ways, as Anderson discusses in his book *More and Different*[19].

Section III suggested that the Hamiltonian in Stern-Gerlach might in theory be expanded to cover the component parts of a silver atom:

$$\hat{H} = \Sigma\, \hat{T}_e + \Sigma\, \hat{T}_u + \Sigma \hat{T}_d + \Sigma\, \hat{V}_{electromagnetic} + \hat{V}_{strong} + \hat{V}_{magnet}$$

This equation covers kinetic and potential energies. $\hat{T}$ are the kinetic energies of the electrons and quarks (up and down). $\hat{V}$ are the potential energies of electromagnetic and strong interactions, plus the external magnetic potential. With such a comprehensive representation, the mass and the magnetic moment would then be observables of the state.

The Transition Recipe does not specify what mathematical methods and what physical theories should be incorporated in a prediction. Richard Feynman in the Character of Physical Law[20] said, "Every theoretical physicist who is any good knows six or seven different theoretical representations for exactly the same physics. He knows that they are all equivalent, and that nobody is ever going to be able to decide which one is right at that level, but he keeps them in his head, hoping that they will give him different ideas for guessing."

## V.    FINAL COMMENTS

The discussion above casts doubts the idea that there is a quantum recipe for making predictions. As Wallace has recently commented[21], the quantum recipe itself makes no predictions. It explains no phenomena. By itself it cannot be tested or falsified.

On a cooking metaphor, for Rule 1 the state is the cooking pot, Rule 2 the interaction is the oven and Rule 3 the measurement is the tasting. The pot plus the oven plus the tasting do not constitute a recipe. There are no ingredients nor any instructions. There is a proverb, "if you like sausages you should never watch them being made." This proverb could well have been about the Stern-Gerlach experiment. In attempting day-to-day calculations, different ingredients are chosen by different physicists, and their utilisation varies from calculation to calculation. Hardly a good omen for philosophers of science seeking rigorous foundations.

Real physics blends ingredients and mixes and matches them and makes progress by breaking rules and guesswork and trial and error. Nor is it a hierarchy with a fundamental basis from which all else follows. As Cartwright suggests[22] it is much more a patchwork, with bits stitched together from different places. The fundamentalist notion that physics follows rigidly from a clearly defined set of immutable laws would preclude any possibility of progress.

Several different representations of Stern-Gerlach are discussed in this paper. It may be tempting to claim they are equivalent - merely translations into different mathematical languages of the same logical argument. This would be wrong. Different formulations may be similar but they produce fractionally different predictions. For that reason, it would be wrong to claim they are equivalent. The vagueness of Rules 1 and 2 has been revealed in the context of Stern-Gerlach.

The fact that the philosophy of science debate focuses on Rule 3 and ignores the other two is suggestive. Maybe to philosophical fundamentalists, as Cartwright calls them, Rules 1 and 2 are vague but acceptably vague.

If there really were a recipe and if physics really was rote learning of recipes, then there would be no need for students to attend physics classes, they could just follow the recipe books, and there would be no need to learn technique or to make trial and error experiments. The same comment could be applied to philosophy. The reality is different. Nobel Prizes are not awarded to those who meticulously follow a recipe.

**Acknowledgements**

I am grateful to David Z Albert, Dustin Lazarovici, Daniel Platt, Christopher Timpson, Bryan Webber and Andrew Whitaker for words of encouragement and helpful comments.